\documentclass[]{aa}
\usepackage{graphicx}
\usepackage[]{natbib}

\bibpunct{(}{)}{;}{a}{}{,} 

\titlerunning{}

\def\tef{\textit{T}_{\text{eff}}}
\def\logg{\text{log}(\textit{g})}
\def\mh{[\text{M}/\text{H}]}

\def\alffe{[\alpha/\text{Fe}]}

\def\ali{\text{A}_{\text{Li}}}

\usepackage{color}
\usepackage{amstext}

\begin{document}

\title{Explaining the decrease in ISM lithium at super-solar metallicities 
in the solar vicinity}

\titlerunning{Explaining the decrease in ISM lithium\\at super-solar metallicities}
\authorrunning{Guiglion et al.}

\author{G. Guiglion \inst{1}, C. Chiappini \inst{1}, D. Romano \inst{2}, 
F. Matteucci \inst{3, 4, 5}, F. Anders \inst{1}, M. Steinmetz \inst{1}, 
I. Minchev \inst{1}, P. de Laverny \inst{6}, A. Recio-Blanco \inst{6}}

\institute{Leibniz-Institut f\"ur Astrophysik Potsdam (AIP), An der Sternwarte 
16, 14482 Potsdam, Germany\\ email: gguiglion@aip.de \and
INAF, Astrophysics and Space Science Observatory, Via Gobetti 93/3, I-40129 
Bologna, Italy \and Dipartimento di Fisica, Sezione di Astronomia, Universit\`a 
di Trieste, via G.B. Tiepolo 11, I-34131, Trieste, Italy \and I.N.A.F.
Osservatorio Astronomico di Trieste, via G.B. Tiepolo 11, I-34131, Trieste, Italy
\and I.N.F.N. Sezione di Trieste, via Valerio 2, 34134 Trieste, Italy 
\and Universit\'e C\^ote d'Azur, Observatoire de la C\^ote d'Azur, CNRS, Laboratoire Lagrange, France}

\date{Received 7 September 2018 / Accepted 7 February 2019}

\abstract{We propose here that the lithium decrease at super-solar metallicities 
observed in high-resolution spectroscopic surveys can be explained by the 
interplay of mixed populations that originate in the inner regions of the Milky 
Way disk. The lower lithium content of these stars is a consequence of inside-out 
disk formation plus radial migration. In this framework, local stars with 
super-solar metallicities would have migrated to the solar vicinity and depleted 
their original lithium during their travel time. To obtain this result, we 
took advantage of the AMBRE catalog of lithium abundances combined with chemical 
evolution models that take into account the contribution to the lithium enrichment 
by different nucleosynthetic sources. A large proportion of migrated stars can explain 
the observed lower lithium abundance at super-solar metallicities. 
We stress that no stellar model is currently able to predict 
Li-depletion for these super-solar metallicity stars, and solar Li-depletion 
has to be assumed. In addition, no solid quantitative estimate 
of the proportion of migrated stars in the solar neighborhood and their travel time is currently available. 
Our results illustrate how important it is to properly include radial migration when
chemical evolution models are compared to observations, and that in this case, the lithium 
decrease at larger metallicities does not necessarily imply that stellar yields have 
to be modified, contrary to previous claims in the literature.}


\maketitle

\section{Introduction}

The chemical evolution of the Milky Way (MW) lithium abundance is still a matter 
of debate. Recent high-resolution spectroscopic surveys have reported that the 
interstellar medium (ISM) lithium content of the MW disk in the solar vicinity 
decreases at super-solar metallicities ($\mh>0.0\,$dex).
\citet{delgado_mena_2015b} were the first to state that lithium tend to decrease for 
$\mh>0$, for which result they used high-resolution spectra. This result has been 
confirmed by \citet{guiglion_2016} using HARPS, FEROS, and UVES spectroscopic data 
from the AMBRE project \citep{laverny_2012}. These data cover a wider range in 
metallicity, up to $\mh\sim0.45\,$dex. The lithium decrease 
at super-solar metallicity was then observed by \citet{fu_2018} using Gaia-ESO 
spectrocsopic survey data and by \citet{bensby_2018} based on high-resolution data from various spectrographs 
(such as MIKE and FEROS).

\begin{figure*}
\centering
\includegraphics[width=1.0\linewidth]{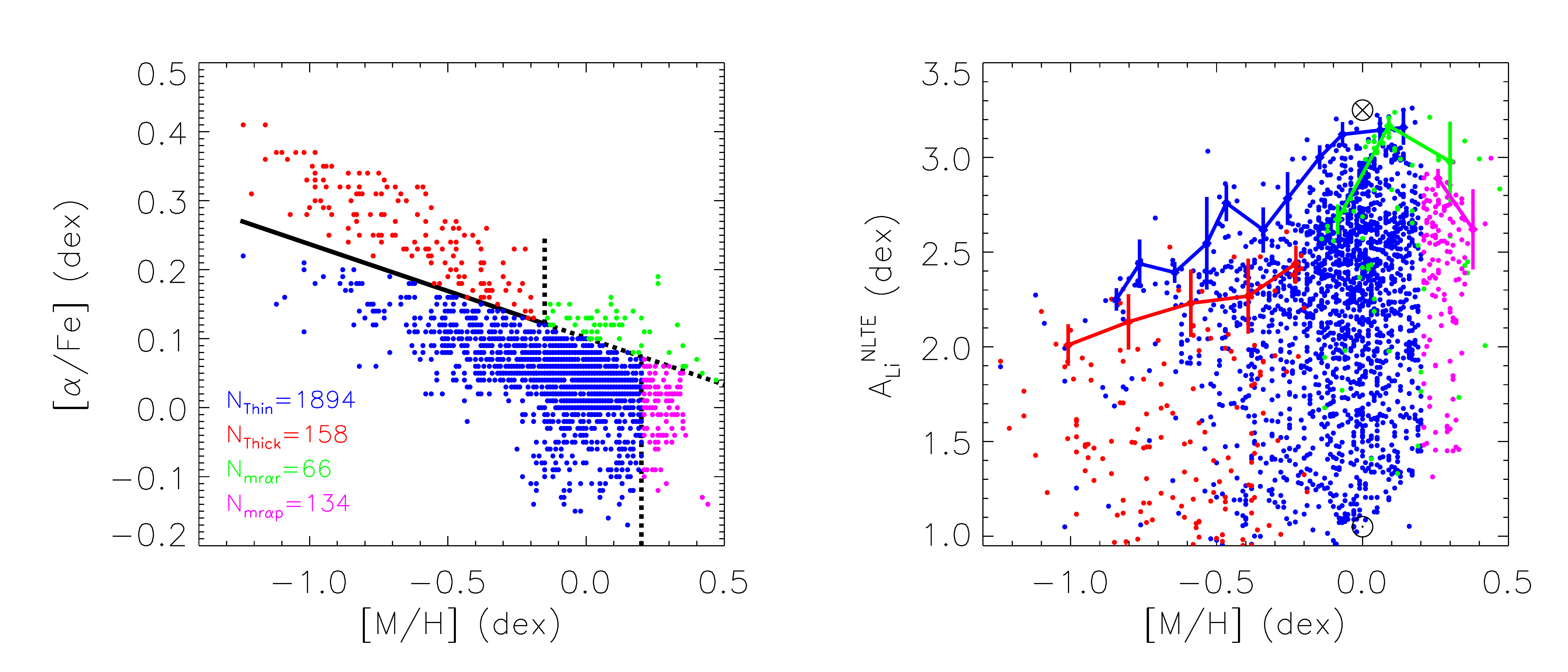}
\caption{\label{alpha_fe_metallicity}\emph{Left}: $\alffe\,vs.\,\mh$ patterns 
for the thin-disk (blue), canonical thick-disk (red), metal-rich $\alffe$-rich 
stars (green), and super-metal-rich solar-$\alpha$ stars (magenta). The black 
lines show how we identified these four MW populations. \emph{Right:} AMBRE NLTE 
lithium abundance $vs.$ metallicity for the corresponding four populations using 
the same color-code as in the left panel. For a given population, we plot the 
maximum lithium abundance using the same method as \citet{guiglion_2016}. The 
meteoric lithium abundance, which is not affected by from stellar evolution, is also shown 
($\otimes$).}
\end{figure*}

This decrease is puzzling because standard chemical evolution models point toward a 
monotonous increase in lithium with metallicity \citep{romano_2001, prantzos_2012}. 
\citet{fu_2018} suggested a possible connection with a reduced nova outburst rate at 
super-solar metallicity. With the \emph{\textup{ad hoc}} assumption of a significant reduction 
of the Li yields above solar metallicity, {\citet{prantzos_2017} reproduced such a decreaes in 
lithium abundance at super-solar metallicity. However, previous stellar 
nucleosynthesis studies pointed out an increase in lithium yields for $\mh>0$ (see, e.g., \citealt{karakas_2016}). 

When we study the Galactic lithium evolution with metallicity, we have to keep in 
mind that lithium is easily depleted in stellar interiors throughout the life of the star. 
Moreover, some lithium can be produced at specific stages of stellar evolution. The
stellar lithium content cannot be assumed to be representative of the Li abundance of the ISM 
material from which the star was formed. Because of the broad spread in lithium for a 
given metallicity, it is important to only consider dwarf stars with the highest lithium, 
corresponding to those that have destroyed at least their photospheric lithium and 
have synthesized none. In practice, when predictions from a Galactic chemical evolution model 
are compared to the observations, the upper envelope of the data 
for dwarf stars needs to be considered (this is believed to be closer to the original lithium in the ISM) as a function 
of metallicity.

In this paper, we provide an explanation for the lithium decrease observed in 
super-solar metallicity thin-disk stars based on the AMBRE lithium catalog of 
\citet{guiglion_2016}, the Galactic chemical evolution model for the thin disk 
of \citet{chiappini_2009_thin_thick}, and our current understanding of the chemodynamic 
evolution of our Galaxy \citep{minchev_2013, minchev_2014b, kubryk_2015}. In 
Sect~\ref{chemical_patterns} we present lithium patterns in different MW disk 
populations, and in Sect~\ref{the_figure} we explain the decrease in lithium boundary 
at high metallicities as due to the presence of old stars ($\tau>5$\,Gyr) that originate in 
the very inner regions that have a  lower lithium content than locally born solar-vicinity stars.

\section{ISM lithium abundance in the MW disk}\label{chemical_patterns}

We used the AMBRE catalog of lithium abundances from \citet{guiglion_2016}. 
This catalog contains chemical abundances of lithium and their non-local thermal equilibrium (NLTE) corrections, derived 
homogeneously from ESO HARPS, FEROS, and UVES spectra. The atmospheric parameters 
have been derived by \citet{worley_2012, worley_2016} and \citet{depascale_2014}, and 
can be found in the publicly available catalog from \citet{guiglion_2018}, with detailed 
chemical abundances of \emph{s-} and \emph{r-}process elements. In this study, we selected 
dwarf stars ($\logg>3.5$) for which lithium abundance and NLTE corrections were available, 
excluding upper limits. We focused on the best stellar parameters and lithium abundances, that is, on
\emph{} stars with AMBRE QUALITY\_FLAG=0 (see \citet{worley_2012} for more details 
on this parameter). We rejected the coolest dwarf stars $(\tef<5\,300\,$K) with a deeper 
convective zone (which results in higher lithium destruction). The median signal-to-noise ratio $(S/N)$ of our 
final sample is 105.

 As recently demonstrated by \citet{anders_2018}, the  use  of  the  dimensionality  
 reduction  algorithm  t-SNE\footnote{t-distributedStochastic Neighbour Embedding 
 \citep{van_der_maaten_2008}} method ( see Section~\ref{the_figure}) allows a more 
 robust definition of subpopulations in abundance space. For this technique to work, 
 it is crucial to confine the analysis to narrow regions in atmospheric-parameter 
 space to avoid spurious abundance trends induced by differences in atmospheric 
 parameters. Although we do not have all the detailed chemical information to be 
 able to compute the t-SNe subpopulations for our sample, we can use those found in \citet{anders_2018} for guidance because the two samples (the sample studied by 
 Anders et al. and ours) are consistent in terms of abundance range and 
  atmospheric parameters (see their Figure 2, using data from \citet{delgado_mena_2017}).

Therefore, we chemically identified the MW disk populations in the 
$\alffe\,\text{versus}\,\mh$ chemical space  \citep[e.g.,][]{fuhrmann_1998, fuhrmann_2011, 
minchev_2015, martig_2016, kawata_2016} using the same definitions as in 
\citet[][see their Fig. 10 and their Sect. 6.2 for more 
details]{guiglion_2016}. We find that the 
thin and chemical thick disks are fully consistent with those of \citet{anders_2018}. 
The main difference is that through the t-SNE method, \citet{anders_2018} also 
identified a metal-rich $\alpha$-rich population (so-called $mr\alpha r$, 
\citealt{adibekyan_2011}), which we here now also include by defining as \textup{\textup{
super-solar metallicity stars}} all thin disk stars with $\mh>0.2\,$dex and 
solar $\alpha$-element abundance. Once again, this is possible because the 
atmospheric parameters and chemical abundances in AMBRE are consistent with the study of 
\citet{delgado_mena_2017}  (see \citet{guiglion_2018} for more details). }
The resulting chemical patterns are presented in the left 
panel of \figurename~\ref{alpha_fe_metallicity}.

After we defined the sub-populations and in order to trace the lithium 
abundance of the ISM as a function of metallicity, we followed the same procedure as 
\citet{lambert_2004} and \citet{guiglion_2016}: we binned the AMBRE data in 
metallicity, and computed an average lithium abundance for the six stars with the 
highest lithium abundance in a given metallicity bin. The NLTE lithium abundance as a 
function of $\mh$ of our sample is shown in the right panel of 
\figurename~\ref{alpha_fe_metallicity}, with different colors corresponding to the 
four populations identified in the left panel. The resulting trends in ISM lithium abundance 
for the four populations are overplotted with the colored lines, with error 
bars corresponding to the standard deviation of the six stars in each metallicity bin. 

The thick-disk lithium envelope first rises from 
$\ali^{\text{NLTE}}\sim2\,$dex at $\mh=-1\,$dex to $\ali^{\text{NLTE}}\sim2.4\,$dex 
at $\mh=-0.2\,$dex. The thin-disk lithium envelope shows an increase of about 1 dex 
and reaches the meteoric abundance at solar metallicity. The envelope for the $mr\alpha r$ 
stars reaches that of the thin disk at solar metallicity and clearly 
decreases by about 0.20 dex at $\mh\sim+0.3\,$dex. In the metal-rich solar-$\alpha$ 
population, the lithium content is clearly lower than in the $mr\alpha r$ 
stars and in the thin-disk stars with solar metallicity. The trends 
presented here remain the same when the slope of the chemical separation is slightly varied
and the bin size of the metallicity axis is changed. In addition, low statistics is 
unlikely to degrade the observed trends because the proportion of stars 
($4\,\%$) with $\ali^{\text{NLTE}}>3.0\,$dex is the same for both $-0.1<\mh<+0.1\,$dex and 
$\mh>+0.1\,$dex. The lithium decrease is therefore not due to a poor sampling of the 
metal-rich populations.

\section{Explaining the lithium decrease at super-solar metallicity}\label{the_figure}

The answer to why in the metal-rich $\alffe$-rich (green) and super-metal-rich 
solar-$\alpha$ (magenta) populations Li decreases with metallicity involves 
three different processes that are simultaneously at work: (1) the Li-enrichment 
in the ISM by different stellar sources \citep[e.g.,][]{prantzos_2017}, (2) the 
Li depletion throughout the evolution of the star \citep[e.g.,][]{lagarde_2018}, and 
(3) the fact that the more metal-rich stars are not the youngest stars, as has 
been shown observationally (e.g.,  \citealt{casagrande_2011, trevisan_2011, 
anders_2017}) and explained theoretically by the process of radial migration (e.g., 
\citealt{roskar_2008, schoenrich_binney_2009, minchev_2013, bird_2013, veraciro_2014, 
matinezmedina_2017}).

In \figurename~\ref{alpha_fe_metallicity_2} we illustrate the points made above and 
investigate the feasibility of our explanation quantitatively as well. Although the 
Li-enrichment is mostly well understood, large uncertainties still prevent a detailed 
quantification of the effects of Li-depletion and radial migration from pure theoretical 
arguments. In the case of the two latter processes, we therefore use empirical data 
to guide our estimates.

\subsection{Lithium enrichment}

\begin{figure*}
\centering
\includegraphics[width=1.0\linewidth]{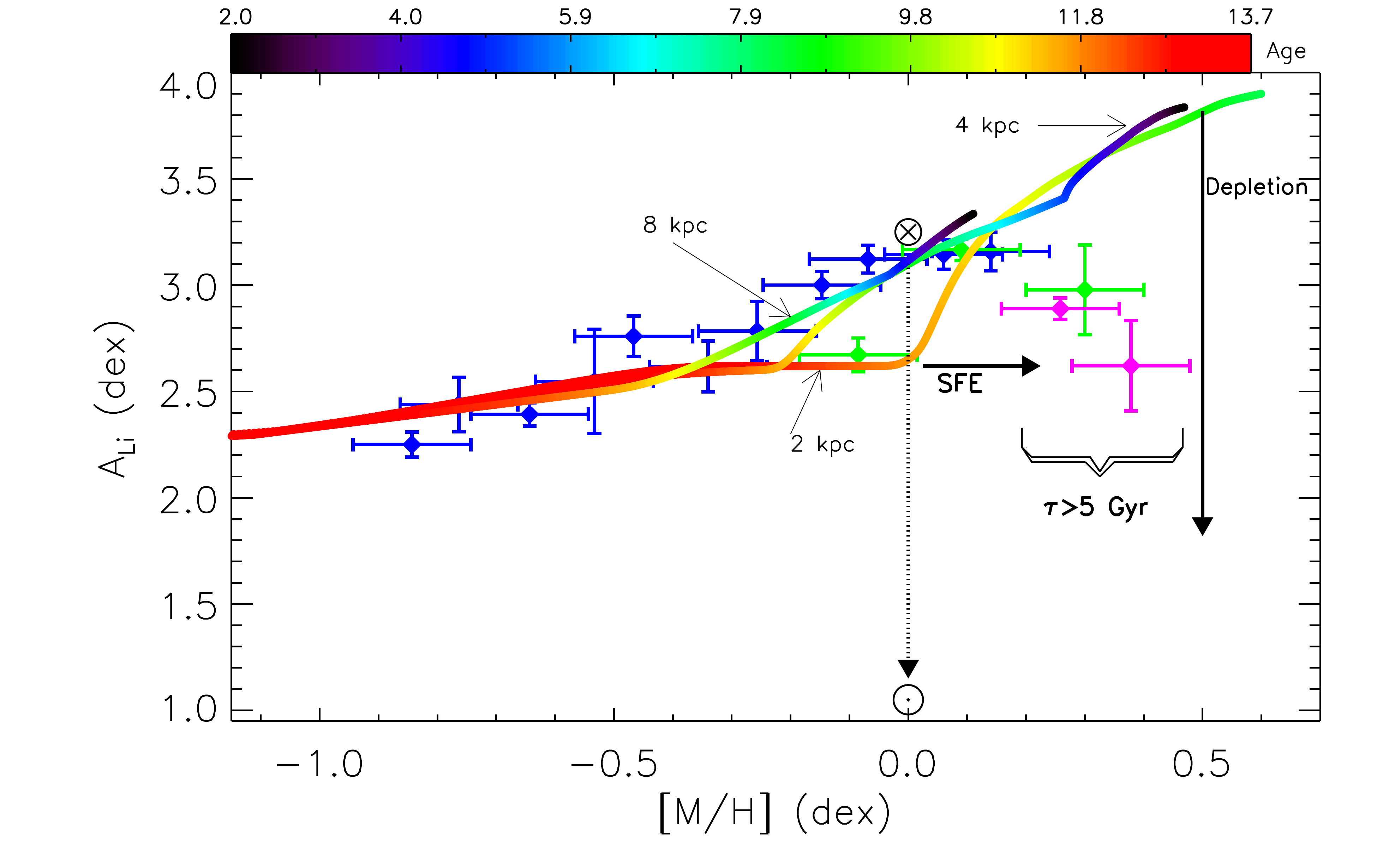}
\caption{\label{alpha_fe_metallicity_2}Lithium abundance $vs.$ 
metallicity for the thin-disk (blue dots), a metal-rich $\alpha-$rich population 
(green dots) and a super-metal-rich solar-$\alpha$ population (pink dots) characterized 
in \figurename~\ref{alpha_fe_metallicity}. We show a Galactic chemical evolution model 
of the thin disk \citep{chiappini_2009_thin_thick}, for three different galactocentric bins of 
2~kpc width (centered at 2, 4, and 8 kpc), color-coded by age. First, inside-out 
chemical evolution models predict the knee of the lithium tracks to occur at higher 
metallicities for the innermost zones because star formation is more efficient (illustrated 
by the horizontal thick arrow). Second, the internal depletion of lithiun in the Sun 
is illustrated by the vertical dashed arrow, from the meteoric abundance to the 
photospheric one. Third, current data suggest super-metal rich stars to be older 
than $\sim5\,$Gyr, which implies a further lithium decrease due to depletion. These old 
ages can be understood in the context of radial migration. However, because there 
is no lithium depletion model at super-solar metallicities, we have assumed this 
depletion to be equal to the solar one (vertical thick arrow).}
\end{figure*}

In \figurename~\ref{alpha_fe_metallicity_2} we show chemical evolution tracks for 
three different radial bins centered on R = 2, 4, and 8 kpc and the data
described in Figure 1. The theoretical curves are color-coded by age, corresponding 
to the color bar. These tracks were computed using the Galactic chemical evolution 
model of \citet{chiappini_2009_thin_thick} (see also \citet{minchev_2013} for more 
details). In this model, the Galactic disk is divided into rings of 2 kpc width that 
evolve independently (\emph{i.e.,} without radial gas flows) in an inside-out fashion 
\citep{matteucci_1989}. This allows us to track the contributors to the ISM lithium 
enrichment with time and radius. The nucleosynthesis prescriptions for lithium adopted 
in the \citet{chiappini_2009_thin_thick} model are the same as in the best model 
of \citet{romano_1999}\footnote{In the model of \citet{romano_1999}, several lithium 
sources in addition to Big Bang nucleosynthesis were considered: cosmic rays impacting 
on ISM nuclei, Type II supernovae (through the $\nu-$ process), carbon stars, massive 
asymptotic giant branch stars (AGBs), and novae. Lithium astration in stars of all masses 
was also taken into account. It was found that a long-lived stellar source was needed 
to reproduce the observations, with novae being recently confirmed as a critical 
ingredient of the model (see \citealt{izzo_2015}).}. We are not 
interested in a detailed discussion of the Li nucleosynthesis here \citep[see][where a discussion can be found]{prantzos_2017, romano_1999, romano_2001}, 
but in showing the general behavior of a chemical evolution model that assumes an 
inside-out formation (i.e., higher star formation efficiencies (SFE) toward the inner 
regions of the Galaxy), and takes into account the several sources of Li nucleosynthesis.

The chemical evolution tracks clearly show that the onset of the strong Li increase, 
mainly powered by the contribution of long-lived sources such as AGBs and novae 
(depending on the details of the chemical evolution model adopted) occurs at higher 
metallicities for the innermost regions\footnote{We here take the 2~kpc curve as our 
innermost example, but more extreme models assuming higher star formation efficiencies 
could be constructed in which the Li knee is displaced to even higher metallicities. 
However, in this case, the final metallicity would be beyond [Fe/H] = 0.6, which is the 
value already reached for the R=2kpc curve, and therefore unrealistic given that no star 
has so far been observed at such a high metallicity.} (a consequence of the higher 
star formation rates; see, e.g., \citealt{matteucci_1990, chiappini_2001, hou_2000, 
minchev_2013}  ; this is illustrated by the horizontal arrow in Figure 2.). \\

\subsection{Lithium depletion throughout the stellar evolution}
 
In addition to lithium evolution in the ISM, lithium is depleted in stellar interiors 
during the main-sequence phase. No stellar model is currently able to reproduce 
the correct lithium depletion in the Sun or predict Li-depletion in [Fe/H]$>$0 
stars. In stellar models that include rotation and thermohaline mixing, the lithium depletion 
does not appear to increase for more metal-rich stars \citep[see the models of][]{lagarde_2018}. 
In Figure 2 we therefore assumed a typical solar depletion for the more metal-rich 
stars in populations green and magenta as well. The arrow in Figure 2 at [Fe/H]$=0.5$ is then not 
arbitrary, but indicates the difference between the pre-solar cloud value measured in 
meteorities and the solar photospheric lithium value (both values are also illustrated 
in Figure~2), which implies a $\sim2.2\,$dex depletion during 4.5 Gyr of evolution 
(illustrated by the dashed arrow at [Fe/H] = 0).

\subsection{Effect of radial migration}

As clearly shown by both observations (e.g., \citealt{casagrande_2011, trevisan_2011, 
anders_2017}) and chemodynamical models (\citet{minchev_2013, minchev_2014}, the most 
metal-rich stars (those with [Fe/H] $>$ 0.25) are typically 5-10 Gyr old.  Therefore 
all of them will have depleted their lithium by large amounts (as discussed in 3.1). 
It follows that the lithium decrease at super-solar metallicities that is observed 
in the solar neighborhood might be explained by radial migration, that is, that 
these stars were not born locally (\citealt{minchev_2013, minchev_2014, minchev_2018, 
anders_2018}, Chiappini et al. 2018 in prep). These stars will have systematically 
lower lithium than young stars that were born at the solar radius (here illustrated by the 
R=8kpc curve).

\section{Conclusion}\label{conclusiooooonnnn}

We have chemically characterized four Milky Way disk populations
using a classical chemical criterion in the well-known $\alffe\,vs\,\mh$ plane: a 
thin-disk, a chemical thick-disk, a metal-rich $\alpha-$rich population, and a 
super-metal-rich solar-$\alpha$ population ($\mh>0.2\,$dex). These populations and 
chemical patterns are consistent with the recent analysis of \citet{anders_2018} and 
Chiappini et al. (in prep). For these four distinct populations, we took advantage 
of the AMBRE catalog of lithium abundances from \citet{guiglion_2016} to trace the 
lithium abundance in the ISM as a function of metallicty. We showed that stars migrating 
from the innermost parts of the Galactic disk (birth radii < 2kpc) will have low 
lithium. If confirmed by stellar age-dating, this finding would rule out the need of 
modifying stellar yields at super-solar metallicity.

This work illustrates the effect that more robust definitions of MW stellar populations 
(allowed by methods such as the t-SNE; \citealt{anders_2018}, Chiappini et al. 
in prep) can have on the recent claim that a Li decline in super-solar metallicity 
stars found by many authors \citep{delgado_mena_2015b, guiglion_2016, fu_2018} would 
imply a revision of the lithium stellar yields at high metallicities. As shown in 
\citet{anders_2018} and Chiappini et al. (in prep), the most metal-rich thin-disk stars 
have most probably migrated from inner disk regions (as has been indicated by previous 
observations such as \citet{casagrande_2011}, \citet{trevisan_2011}, and 
\citet{anders_2017b}; see the discussion in \citealt{chiappini_2015}), and cannot be 
modeled as part of a local thin disk without taking into account the fact that stars 
move out of their birth places, as is clearly shown by chemodynamical models 
(\citealt{minchev_2013, minchev_2014} and \citealt{minchev_2018}). It is also consistent 
with the kinematics found in super-solar metallicity stars in the solar neighborhood, see, 
\emph{e.g.,} \citet{wojno_2018}.

\begin{acknowledgements}
The authors acknowledge the anonymous referee for the comments and suggestions 
that improved the readability of the paper. GG and CC thank Corinne Charbonnel 
for useful discussions. CC acknowledge the DFG project number 283705981: 
"Analysing the chemical fingerprints left by the first stars: chemical abundances 
in the oldest stars". ARB acknowledges financial support form the ANR 14-CE33-014-01. 
IM acknowledges support by the Deutsche Forschungsgemeinschaft under the grant MI 2009/1-1.
\end{acknowledgements}

\bibliographystyle{aa}
\bibliography{cite_r_s}

\end{document}